\documentclass[twocolumn,showpacs,preprintnumbers,amsmath,amssymb]{revtex4}
\input psfig.sty
\usepackage{graphicx}% Include figure files
\usepackage{dcolumn}% Align table columns on decimal point
\usepackage{bm}% bold math

\begin{document}

\title{Non-extensive statistics and the stellar polytrope index}

\author{R. Silva} \email{rsilva@uern.br}
\affiliation{Departamento de F\'{\i}sica, Universidade do Estado
do Rio Grande do Norte, 59610-210, Mossor\'o - RN, Brasil}

\author{J. S. Alcaniz} \email{alcaniz@dfte.ufrn.br}
\affiliation{Departamento de F\'{\i}sica, Universidade Federal do
Rio Grande do Norte, C.P. 1641, Natal - RN, 59072-970, Brasil}

\date{\today}% It is always \today, today,

\begin{abstract}
We use physical constrains imposed from the H-Theorem and from the negative nature of
the heat capacity of self-gravitating thermodynamically isolated systems to investigate
some possible limits on the stellar polytrope index $n$ within the domain of a
classical non-extensive kinetic theory.
\end{abstract}

\maketitle

Polytropic distribution functions provide the simplest models for
self-gravitating stellar systems \cite{binney94}. These models are
characterized by an equation of the state of the form $p=K
\rho^{\gamma}$, where $K$ and $\gamma$ are constants. The
polytropic index $n$ is defined by the following expression:
$\gamma = 1 + {1\over n}$ which in the limit $n\rightarrow \infty$
or, equivalently, $\gamma =1$ results in an equation of state
identical to the one of an isothermal body of gases, i.e., $p=K
\rho$. The distribution function associated to these systems is
given by \cite{binney94}
\begin{equation}
f(\epsilon)={\rho_1\over (2\pi\sigma^2)^{3/2}}e^{\epsilon
/\sigma^2}
\end{equation}
where $\rho_1$ and $\sigma=(k_B T/m)^{1/2}$ are, respectively, the
density and the velocity dispersion. The quantity $\epsilon$ is
equal to a constant minus the total energy per unit mass of an
individual star moving in the global galactic gravitational
potential. In fact, $\epsilon$ is the relative energy of a star
defined by $\epsilon=\Psi-v^2/2$, where $\Psi$ is the relative
potential, given by $\Psi=\phi_0-\phi$, $\phi$ is the star
potential energy and $\phi_0$ is a arbitrary constant. In
particular, the integrability condition $\int f d^3 {\bf
v}<\infty$ results in the constraint $n>1/2$ (for a complete study
on the gaseous polytropes and isothermal gas spheres, see
\cite{chandra39}).

On the other hand, a non-extensive statistical formalism has been
proposed as a possible extension of the classical one, which
intends to study more properly systems that possess long-range
interactions. Such a formalism is based on the entropy formula
\cite{const}
\begin{equation}
S_q=-k_B\sum_i p_i^q \ln_q p_i,
\end{equation}
where $p_i$ is the probability of the ith microstate and $q$ is
the non-extensive parameter. The Botzmann-Gibbs extensive formula,
$S_1=-k_B\sum_i p_i\ln p_i$, is readily recovered in the extensive
limit $q=1$ and the $q$-logarithmic function above is defined as
\cite{tsallisnet}
\begin{equation}
\ln_q f ={f^{1-q}-1\over 1-q},\quad f>0.
\end{equation}

Some attempts to build a kinetic counterpart for this
non-extensive statistical formalism has been recently proposed in
Refs. \cite{rai,lima1}. The main result of these works consists in
showing that this extended formulation leads to a new velocity
distribution for free particles given by
\begin{equation}\label{fq}
f_0(v)=B_q \left[1-(1-q){mv^2\over 2k_B T}\right]^{1/1-q},
\end{equation}
where the quantity $B_q$ is a $q$-dependent normalization constant
whose expressions for $ q \leq 1$ and $ q \geq 1$ are respectively
given by
\begin{subequations}
\begin{equation}
B_{q \leq 1} = {\cal{G}}\frac{\Gamma (\frac{1}{2}+{1\over
1-q})}{\Gamma({1\over 1-q})}\left(\frac{m}{2\pi k_BT}\right)^{3/2}
\end{equation}
and
\begin{equation}
B_{q \geq 1} = {\cal{H}}\frac{\Gamma({1\over q-1})}{\Gamma
({1\over q-1}-\frac{3}{2})}\left(\frac{m}{2\pi k_BT}\right)^{3/2}
\end{equation}
\end{subequations}
where the coefficients ${\cal{G}}$ and ${\cal{H}}$ are given by
${\cal{G}} = n_p(1-q)^{1/2}(\frac{5-3q}{2})(\frac{3-q}{2})$ and
${\cal{H}} = n_p(q-1)^{3/2}$, $T$ is the temperature and $n_p$ is
the particle number density. This $q$-distribution can be derived
at least from two different methods, namely, a generalization of
the Maxwell ansatz, $f(v)\neq f(v_x)f(v_y)f(v_z)$, which follows
from the introduction of statistical correlations between the
components of the velocities \cite{rai} and a new formulation for
the Boltzmann $H$-theorem, which requires $q > 0$ \cite{lima1}.

In the astrophysical domain, both the non-extensive statistical
formalism and its kinetic counterpart have been applied in a wide
range of problems. The very first application of this
non-extensive statistics was done in connection with stellar
polytropes \cite{plast}, with several authors suggesting different
expressions for the polytropic index $n$ as a function of the
non-extensive parameter $q$
\cite{taruya03a,taruya03b,plast01,lima03}. Recently, a numerical
study of stellar dynamical evolution for self-gravitating systems
was performed by Taruya and Sakagami \cite{taruyaPRL03} while
Chavanis \cite{chavanis} considered the Tsallis' entropy as a
particular $H$-function corresponding to isothermal stellar
systems and stellar polytropes. In this latter approach, the
maximization of the $H$-function at fixed mass and energy reveals
a thermodynamical analogy with the study of the dynamical
stability in collisionless stellar systems. The non-extensive
kinetic approach has been used to study the Jeans' gravitational
instability \cite{lima,du}.

Another recent application of this non-extensive kinetic theory  was to investigate the
negative nature of the heat capacity for isolated self-gravitating system
\cite{ra2003}. This study was performed by considering the following steps: first,
consider a simple analogy between
self-gravitating system and ideal gas (IG) or, in other words, the
equivalence between IG kinetical and internal energies, given by
${1\over 2}m<v^2>={3\over 2} k_B T$, where $m$ is the mass of a
particle (e.g., a star). If such a system is composed by $N$
particles, its total kinetic energy is $K={3\over 2}Nk_BT$ which,
according to the viral theorem, is equal to minus total energy,
i.e., $E=-K$. Therefore, the heat capacity of the system is given
by (see \cite{binney94} for details)
\begin{equation}
C_V={dE\over dT}=-{3\over 2}Nk_B.
\end{equation}
Second, in order to investigate
the $q$-dependence of the heat capacity, we consider a cloud of
ideal gas within the non-relativistic gravitational context, which
is analog to a self-gravitating collisionless gas. The kinetic
energy of this system is simply given by $K={1\over 2}m<v^2>$,
with the statistical content included in the average square
velocity of the particles $<v^2>$. Indeed, the non-extensivity may
be introduced through a new derivation of expectation value \cite{tsallis98}
\begin{equation}
<v^2>_q ={\int_{-v_m}^{v_m}f^q v^2 d^3v\over\int_{-v_m}^{v_m}f^q
d^3v},
\end{equation}
where $v_m=\left({2k_B T\over m(1-q)}\right)^{1/2}$ is a thermal
cutoff on the maximum value allowed for the velocity of the
particles $(q<1)$, whereas for the power law without cutoff
$(q>1)$ $v_m\rightarrow\infty$. This $q$-expectaion value can be
easily evaluated for $q\neq 1$, resulting in
\begin{equation}
<v^2>_q={6\over 5-3q} {k_B T\over m},\quad {\rm for} \quad q<5/3.
\end{equation}
 Now, combining
the $q$-expectation value for the square velocity of the particles
with the definition of the heat capacity $C_V$ one finds
\begin{equation}\label{CV<0}
C_V=-{3\over 5-3q} Nk_B,
\end{equation}
which clearly places an upper limit to the non-extensive
parameter, i.e., $q<5/3$. Following this reasoning, it is
therefore natural that our kinetic approach uses the free particle
distribution, with the long range nature of gravity being
introduced via the virial theorem (see \cite{ra2003} for details).
We note that, in the domain of ensemble theory, there are some
analyses in the literature dealing with the classical IG within
this nonextensive scenario
\cite{plastino94,Abe99a,Abe99b,Botelho03}. In particular, Ref.
\cite{Abe99a} showed that the IG naturally exhibits negative heat
capacity and that its pressure and particle number density obey a
polytrope-type relation. Here, however, we assume that all IG
properties are accounted for by the $q$-Maxwellian (see
\cite{rai,ra2003} for details).

In this {\it Letter}, by considering the different expressions
relating the polytropic index $n$ and the non-extensive parameter
$q$ we discuss some possible physical constraints on $n$ which
arise from the second law of thermodynamics ($q > 0$) and the
negativeness of the heat capacity ($q < 5/3$). We show that the
existent expressions between $n$ and $q$ (see
\cite{taruya03b,plast01,lima03}) can be naturally linked with the
heat capacity imposing limits on the polytropic index. It is worth mentioning that the
constraint $q > 0$, arising from the generalized $H$-theorem, does not provide a
strong case for limiting the polytropic index $n$. This corresponds to a weak condition
of the generalized $H$-theorem since it is possible to show that there are two arrow of
time for $q>0$ and $q<0$.

\begin{figure}[t]
\vspace{.2in}
\centerline{\psfig{figure=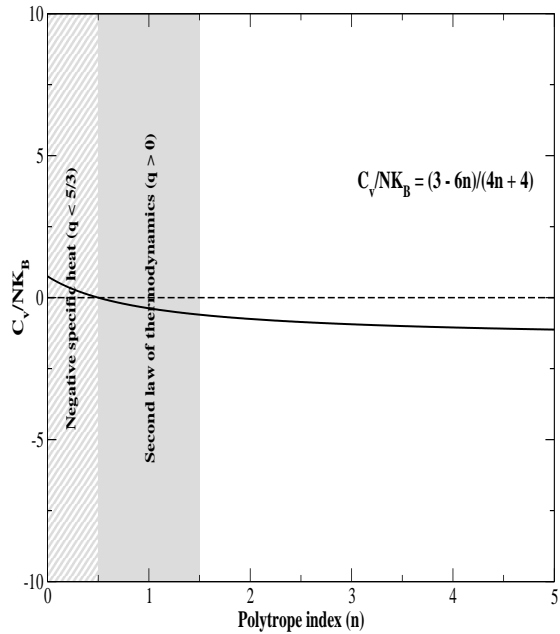,width=3.0truein,angle=-90
,height=3.5truein}\hskip 0.1in} \caption{The quantity $C_V/Nk_B$
for a self-gravitating collisionless gas obeying the non-extensive
Tsallis' $q$-statistic is shown as a function of the polytropic
index $n$. The polytropic index and non-extensive parameter $q$
are related by Eq. (\ref{cv2}). As indicated, the shadowed regions
stand for the constraints from the second law of thermodynamics
and the negative nature of $C_V$.}
\end{figure}

In order to investigate such constraints we first present the
different expressions for the  polytropic index $n$ as a function
of the non-extensive parameter $q$ discussed in the literature,
i.e.,

\begin{equation}\label{tar03}
\mbox{(i)} \quad   \quad   \quad   n={1\over 2}+ {1\over 1-q}\quad [8],
\end{equation}
and
\begin{equation}\label{lima}
\mbox{(ii)} \quad   \quad   \quad   n={5-3q\over2(1-q)}\quad [9,10],
\end{equation}

\begin{figure}[t]
\vspace{.1in}
\centerline{\psfig{figure=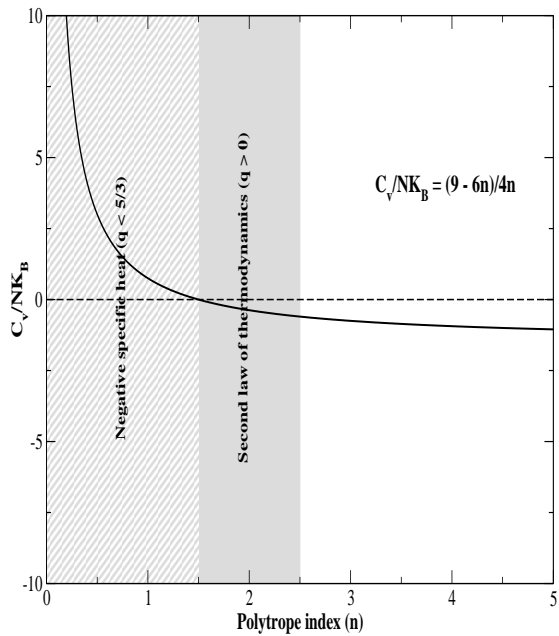,width=3.0truein,angle=-90
,height=3.5truein}\hskip 0.1in} \caption{The same as in Fig. 1 for
the case (ii). Here the polytropic index and non-extensive
parameter $q$ are related by Eq. (\ref{cv3}).}
\end{figure}

The case (i) arises from an extension of Padmanabhan's classical
analysis of the Antonov instability for the case of polytropic
distribution function \cite{pad}. By using kinetic arguments of
the Tsallis' non-extensive formalism, the case (ii) is
obtained from an extension of the polytropic Lane-Emden spheres as
well as from a study of maximum entropy solutions of the
Vlasov-Poisson equations describing self-gravitating systems. We emphasize that the
polytropic family of solutions to the collisionless Boltzmann equation (parameterized
by the index $n$) is always the same, which means that expressions (\ref{tar03}) and
(\ref{lima}) constitute only different ways of parameterizing (in
terms of non-extensive parameter $q$) this polytropic family. In both cases, Eqs.
(\ref{tar03}) and
(\ref{lima}), the same choice of
the statistical average, the so-called $q$-expectation value, has been used. As
expected, for all the above expressions the Maxwellian isothermal
spheres are obtained in the limit $q=1$ or, equivalently,
$n=\infty$.

By combining Eqs. (\ref{tar03})-(\ref{lima}) with Eq.
(\ref{CV<0}), one obtains the following expressions relating the
heat capacity and the polytropic index:

\begin{equation}\label{cv2}
\mbox{(i)} \quad   \quad   \quad   C_V={(3-6n)\over 4n+4}Nk_B,
\end{equation}
and
\begin{equation}\label{cv3}
\mbox{(ii)} \quad   \quad   \quad   C_V={(9-6n)\over 4n}Nk_B.
\end{equation}

\begin{table}[t]
\caption{Limits to $n$}
\begin{ruledtabular}
\begin{tabular}{lcl}
Thermodynamic constraints& case (i) & case (ii) \\
\hline \hline \\
$q > 0$ & ${1\over 2}<n<{3\over 2}$& ${3\over 2} < n < {5\over
2}$ \\
$q < 5/3$ & ${0} < n < {1\over 2}$ & ${0}<n<{3\over 2}$\\
\end{tabular}
\end{ruledtabular}
\end{table}

From the above expressions, it is straightforward to see that by
imposing the constraints on the non-extensive parameter $q$ from
the second law of thermodynamics ($q > 0$) and the negative nature
of $C_V$ ($q < 5/3$) it is possible to limit regions in the $C_V -
n$ plane. To better visualize such constraints, in Figs. 1 and 2
we show, respectively for the cases (i) and (ii), the
dimensionless quantity $C_V/Nk_B$ as a function of the polytropic
index $n$.

In Figures 1 [case (i)] and 2 [case (ii)] we show that the
constraint $q > 0$ results, respectively, in the forbidden
intervals for the polytropic index ${1\over 2}<n<{3\over 2}$ and
${3\over 2} < n < {5\over 2}$ while the limit resulting from Eq.
(\ref{CV<0}), i.e., $q < {5\over 3}$ implies ${0} < n < {1\over
2}$ and ${0}<n<{3\over 2}$. In all cases the bound on $q$ from the
second law of thermodynamics ($q > 0$) restricts only negative
regions in the plane $C_V/Nk_B - n$ while, as expected, the
themodynamical limit $C_V < 0$ forbids any possible positive
region in the plots. In particular, we also note that the
constraints on $n$ from Eq. (\ref{cv2}) (case i) seem to be in
agreement with the theorem proved by Antonov and Dremus
\cite{binney94} who showed that the equilibrium distribution is
dynamically stable for values of the polytrope index $n > 3/2$ or,
equivalently,  $q
> 0$ (see Fig. 1). The bounds on the polytopic index $n$ obtained in this \emph{Letter}
are summarized in Table I.

Finally, we emphasize that in a more quantitative case, i.e., in
which a self-gravitating ideal gas is contained in a spherical
container of radius $R$ our constraints on the polytropic index
$n$ must be valid only in the negative region of the heat
capacity. As widely known, when the potential energy is taking
into account or, equivalently, the virial theorem is modified to
introduce a pressure term, $2K + W \simeq p(R)$, the heat capacity
can assume both positive and negative values. We also note that a
possible connection with stability of spherical systems can be
introduced through the Doremus-Feix-Baumann theorem which implies,
for the extensive case, that polytropes with ${1\over 2}<n<{3\over
2}$ are unrealistic (see, however, \cite{henon73,barnes86}).

\begin{acknowledgments}
The authors are very grateful to A. Taruya, J. R. da Silva and
Prof. J. A. S. Lima for helpful discussions and critical reading
of the manuscript. We also thank the anonymous referee for
valuable comments. JSA is supported by the Conselho Nacional de
Desenvolvimento Cient\'{\i}fico e Tecnol\'{o}gico (CNPq - Brasil)
and CNPq (62.0053/01-1-PADCT III/Milenio).
\end{acknowledgments}


\begin{thebibliography}{30}

\bibitem{binney94} J. Binney, S. Tramaine {\it Galactic Dynamics}
(Princeton U. Press, 1994).

\bibitem{chandra39} S. Chandrasekhar {\it An introduction to the
Theory of Stellar Structure} (University of Chicago Press, 1958).

\bibitem{const} C. Tsallis, J. Stat. Phys. {\bf 52} (1988) 479.

\bibitem{tsallisnet} See also
http://tsallis.cat.cbpf.br/biblio.htm for an up to date
bibliography.

\bibitem{rai} R. Silva, A. R. Plastino, J. A. S. Lima, Phys.
Lett. A {\bf 249} (1998) 401.

\bibitem{lima1} J. A. S. Lima, R. Silva, A. R. Plastino, Phys. Rev.
Lett. {\bf 86} (2001) 2938. cond-mat/0210072

\bibitem{chavanis} P. H. Chavanis,  A\&A {\bf 386} (2002) 732; {\bf
401} (2003) 15.


\bibitem{taruya03b} A. Taruya, M. Sakagami, Physica A {\bf 322} (2003)
285.

\bibitem{plast01} A. R. Plastino, in {\it Nonextensive Statistical
Mechanics
and Its Aplications, eds. S. Abe and Y. Okamoto} (Springer,
Berlin, 2001)

\bibitem{lima03} J. A. S. Lima, R. E. de Souza, {\it On the
Nonextensive Isothermal Stellar Spheres}. Submitted for
publication.

\bibitem{plast} A. Plastino, A. R. Plastino, Phys.
Lett. A {\bf 174} (1993) 384.

\bibitem{taruya03a} A. Taruya, M. Sakagami, Physica A {\bf 318} (2003) 387

\bibitem{taruyaPRL03} A. Taruya, M. Sakagami, Phys. Rev. Lett. {\bf 90} (2003) 181101.

\bibitem{lima} J. A. S. Lima, R. Silva, J. Santos, A\&A {\bf 396}
(2002) 309. astro-ph/0109474

\bibitem{du} J. L. Du, Phys. Lett. A {\bf{320}} (2004) 347

\bibitem{ra2003} R. Silva, and J. S. Alcaniz, Phys. Lett. A {\bf 313},
(2003) 393. astro-ph/0306009

\bibitem{plastino94} A. R. Plastino, A. Plastino and C. Tsallis,
J. Phys. A {\bf 27}, (1994) 5707.


\bibitem{Abe99a} S. Abe, Phys. Lett. A {\bf 263}, (1999) 424.

\bibitem{Abe99b} S. Abe, Physica A {\bf 269}, (1999) 403.

\bibitem{Botelho03} L. C. L. Botelho, Mod. Phys. Lett. B {\bf 17}, (2003) 733.

\bibitem{tsallis98} C. Tsallis, R. S. Mendes and A. Plastino, Physica A {\bf 261} (1998)
534.

\bibitem{pad} T. Padmanabhan, Phys. Rep. {\bf 188} (1990) 285.

\bibitem{henon73} M. Henon,  A\&A {\bf 24} (1973) 229.

\bibitem{barnes86} J. Barnes, P. Hut, Nature {\bf 324} (1986) 446.




\end{thebibliography}

\end{document}